\documentclass[12pt]{article}
\usepackage{amssymb}
\usepackage{amsmath}

\begin{document}
\begin{center}
\vspace*{2cm} {\Large {\bf KPZ Scaling Theory and the
Semi-discrete Directed Polymer Model \bigskip\bigskip\\}}
{\large{Herbert Spohn}}\bigskip\bigskip\\
   {Department of Mathematics and Physics, TU M\"unchen,\\
 D-85747 Garching, Germany\\e-mail:~{\tt spohn@ma.tum.de}}\\

\end{center}
\vspace{5cm} \textbf{Abstract.} We explain how the claims of the KPZ scaling theory 
are confirmed by a recent proof of 
Borodin and Corwin on the asymptotics of the semi-discrete directed 
polymer.

\newpage

\section{Introduction}

The 1986 Kardar-Parisi-Zhang (KPZ) equation \cite{KPZ} is a stochastic
partial differential equation modeling surface growth and, more
generally, the motion of an interface bordering a stable against a
metastable phase. In a landmark contribution, Krug,
Meakin, and Halpin-Healy \cite{KMH} developed a scaling theory, which is an educated guess on
the non-universal coefficients in the asymptotics for
models in the KPZ universality class. 
 The purpose of our note is to present the scaling theory in a modernized form, in view of the 
 much advanced understanding, and 
 to explain how to apply the
scaling theory to the semi-discrete directed polymer. This model has
been discussed in depth at the 2010 random matrix workshop at the
MSRI and, so-to-speak as a spin-off, Borodin and Corwin \cite{BC} developed the
beautiful theory of Macdonald processes, which provides the tools
for an asymptotic analysis of the semi-discrete directed polymer model. As
we will establish, the scaling theory is consistent with the results in
\cite{BC,BCF}, thereby providing a highly non-obvious control check.

To place the issue in focus, let me start with a simple example.
Assume as given the stationary sequence $X_j$, $j\in\mathbb{Z}$, of 
random variables and let us consider the partial sums
\begin{equation}\label{1}
S_n=\sum^{n}_{j=1} X_j\,.
\end{equation}
As well studied, it is fairly common that $\big(S_n - \mathbb{E}(X_0)n\big)/\sqrt{n}$ converges
to a Gaussian as $n\to \infty$, i.e.
\begin{equation}\label{2}
\lim_{n\to\infty} \mathbb{P}\big(S_n - \mathbb{E}(X_0)n \leq \sqrt{D}\sqrt{n}
s\big)=F_\mathrm{G}(s)\,,
\end{equation}
where $F_\mathrm{G}$ is the distribution function of a unit Gaussian
random variable. Here $F_\mathrm{G}$ is the \textit{universal} object, while $\mathbb{E}(X_0)$ and
the coefficient $D>0$ depend on the law $\mathbb{P}$ and are in this
sense model dependent, in other words \textit{non-universal}. Using
stationarity, $D$ is readily guessed as
\begin{equation}\label{3}
D=\sum^{\infty}_{j=-\infty}\big( \mathbb{E}(X_0 X_j) - \mathbb{E}(X_0)^2 \big)\,.
\end{equation}

The KPZ class deals with strongly dependent random variables, for
which partial sums are of size $n^{1/3}$ rather than $n^{1/2}$.
$F_\mathrm{G}$ is to be substituted by the GUE Tracy-Widom
distribution function, $F_\textrm{GUE}$, which first appeared in the
context of the largest eigenvalue of a GUE random matrix \cite{F93,TW95}.
$F_\textrm{GUE}$ is defined through a Fredholm determinant as
\begin{equation}\label{4}
F_\textrm{GUE}(s)= \det (1-P_s K_\textrm{Ai} P_s)\,.
\end{equation}
Here $K_\textrm{Ai}$ is the Airy kernel,
\begin{equation}\label{5}
K_\textrm{Ai}(x,y)=\int^{\infty}_0 d\lambda\, \textrm{Ai}(x+\lambda)
\textrm{Ai}(y+\lambda)\,,
\end{equation}
with Ai the Airy function, and $P_x$ projects onto the interval
$[x,\infty)$. For each $x$, $P_x K_\textrm{Ai} P_x$ is a trace class
operator in $L^2(\mathbb{R})$, hence (\ref{4}) is well-defined. To
determine the scale coefficient is less obvious than in the example
above, but will be explained in due course. Let me stress that the
scaling theory is crucial for the proper statistical analysis of
either physics \cite{KS,KSSS} or computer experiments \cite{OFA}. Without this
input, the comparison with theoretical results would be considerably
less reliable.

Our paper is divided into two parts. We first explain the scaling
theory in the context of a specific class of growth models. In the
second part the theory is applied to the semi-discrete directed
polymer model. The convergence to the GUE Tracy-Widom distribution
is established in \cite{BC,BCF}, of course including an expression for the
non-universal scale coefficient. Our goal is to explain, how this
coefficient can be determined independently, not using the import
from the proof in \cite{BC,BCF}.


\section{Scaling theory for the single-step model}\label{sec2}
 \setcounter{equation}{0}

In the single step growth model the moving surface is described by
the graph of the height function $h(t):\mathbb{Z}\to\mathbb{Z}$ with
the constraint
\begin{equation}\label{6}
|h(j+1,t)-h(j,t)|=1\,,
\end{equation}
hence the name. The random deposition/evaporation events are modeled
by a Markov jump process constrained to satisfy (\ref{6}). The
allowed local moves are then transitions from $h(j,t)$ to $h(j,t)\pm
2$. The dynamics should be invariant under a shift in the
$h$-direction. Hence the rates for the deposition/evaporation events
are allowed to  depend only on the local slopes. It is then
convenient to switch to height differences
\begin{equation}\label{7}
\eta_j(t)=h(j+1,t)-h(j,t)\,,\quad \eta_j(t)=\pm 1\,.
\end{equation}
A single growth step  at the bond $(j,j+1)$ is given by
\begin{equation}\label{8}
\eta \rightarrow \eta^{j,j+1}\,,
\end{equation}
where $\eta^{j,j+1}$ is the configuration with the slopes at $j$ and
$j+1$ interchanged. If the corresponding rates are denoted by
$c_{j,j+1}(\eta)$, depending on the local neighborhood of $(j,j+1)$,
the Markov generator reads
\begin{equation}\label{9}
L f(\eta)=\sum_{j\in\mathbb{Z}} c_{j,j+1}(\eta)
\big(f(\eta^{j,j+1})- f(\eta)\big)\,.
\end{equation}
The slope field $\eta$ is locally conserved, in the sense that the
sum $\sum^b_{j=a} \eta (j,t)$ changes only through the fluxes at the two
boundaries $a$, $b$. To keep things concretely, in the following we
consider only the wedge initial condition
\begin{equation}\label{10}
    h(j,0)=|j|\,.
\end{equation}

The scaling theory is based on the\medskip\\
\textbf{Assumption.} \textit{The spatially ergodic and time stationary measures of the slope process $\eta(t)$ are
precisely labeled by the average density
\begin{equation}\label{11}
\rho=\lim_{a\to \infty} \frac{1}{2a+1} \sum_{|j|\leq a} \eta_j
\end{equation}
with $|\rho|\leq 1$.}\medskip\\
Our assumption has been formulated more than 30 years ago. Except
for special cases, it remains open even today, see the book by
Liggett \cite{L98} for more details. The stationary measures from
the assumption are denoted by $\mu_\rho$, as a probability measure on
$\{-1,1\}^\mathbb{Z}$.

Given $\mu_\rho$ one defines two natural quantities:\medskip\\
$\bullet$ the average steady state current
\begin{equation}\label{12}
{\sf j}(\rho)=\mu_\rho (c_{0,1}(\eta)(\eta_0-\eta_1))
\end{equation}
$\bullet$ the integrated covariance of the conserved slope field
\begin{equation}\label{13}
A(\rho)=\sum_{j\in\mathbb{Z}}\big(\mu_\rho
(\eta_0\eta_j)-\mu_{\rho}(\eta_0)^2\big)\,.
\end{equation}

We first notice that for long times there is a law of large numbers
stating that
\begin{equation}\label{14}
h(j,t)\simeq t \phi(j/t)
\end{equation}
for large $j,t$ with a deterministic profile function $\phi$. In
fact $\phi$ is the Legendre transform of ${\sf j}$ in the sense that
\begin{equation}\label{15}
\phi (y)= \sup_{|\rho|\leq 1} \big(y\rho-{\sf j}(\rho)\big)\,.
\end{equation}
The argument is based on the hydrodynamic limit for nonreversible
lattice gases \cite{S91}, which asserts that on the macroscopic scale
the density $\rho(x,t)$ of the conserved field satisfies
\begin{equation}\label{16}
\frac{\partial}{\partial t}\rho (x,t)+\frac{\partial}{\partial
x}{\sf j}(\rho (x,t))=0\,
\end{equation}
with initial condition
\begin{equation}\label{17}
\rho(x,0)=\left\{
            \begin{array}{ll}
              1\,, & {x\geq 0,} \\
              -1\,, & {x<0.}
            \end{array}
          \right.
\end{equation}
The entropy solution to (\ref{16}), (\ref{17}) is indeed given by
(\ref{14}), (\ref{15}).

The average current can be fairly arbitrary, except for the linear behavior near $\rho = \pm 1$. On the other hand $\phi$ is convex up with $\phi(x) = |x|$ for $|x| \geq x_\mathrm{c}$. At points, where $\phi$  is either  linear or cusp-like, the fluctuations may be different from the generic case and have to be discussed separately.
We set 
\begin{equation}\label{17a}
\lambda(\rho) = - {\sf j}''(\rho)\,.\medskip
\end{equation}
\textbf{Conjecture} (KPZ class). \textit{Let $y$ be such that $\phi$ is
twice differentiable at $y$ with $\phi''(y)\neq 0$ and set
$\rho=\phi'(y)$, $|\rho|<1$. If $A(\rho)<\infty$ and
$\lambda(\rho)\neq 0$, then}
\begin{equation}\label{18}
\lim_{t\to\infty} \mathbb{P}\big(h(\lfloor y t\rfloor,t)-t\phi
(y)\leq -(-\tfrac{1}{2}\lambda A^2)^{1/3} t^{1/3}
s\big)=F_{\mathrm{GUE}}(s)\,.\medskip
\end{equation}
$\lfloor\cdot\rfloor$ denotes integer part. Since $\phi''(y)>0$, because of Legendre transform $\lambda(\rho) < 0$. 
 
 On the
scale $(-\tfrac{1}{2}\lambda A^2 t)^{1/3}$ the height fluctuations
are governed by the Tracy-Widom distribution.
$\lambda A^2$ is the model dependent coefficient which, at least in
principle, can be computed once $\mu_\rho$ is available.
$\lambda\equiv 0$ for reversible slope dynamics, since ${\sf j}=0$. In
that case the fluctuations are of scale $t^{1/4}$ and Gaussian, see
\cite{S91}, Part II, Chapter 3, for a discussion. For nonreversible slope dynamics one
can still arrange for $\lambda\equiv 0$. If ${\sf j}$ is non-zero,  there could be isolated points at which
$\lambda$ vanishes. At a cubic inflection point
of ${\sf j}$ the height fluctuations are expected to be of order $(t(\log t)^{1/2})^{1/4}$
\cite{P,DLSS}.

In our set-up the conjecture has been proved by Tracy and Widom
\cite{TW09} for the PASEP. In this case the exchange $+-$ to $-+$
occurs with rate $p$ and the exchange $-+$ to
$+-$ with rate $1-p$, $ 0 \leq p < \frac{1}{2}$. Then $\mu_\rho$ is a Bernoulli measure, hence
$A(\rho)=1-\rho^2$, ${\sf j}(\rho)= \frac{1}{2}(2p - 1)(1-\rho^2)$, and
\begin{equation}\label{19}
- \tfrac{1}{2}\lambda A^2= \tfrac{1}{2}(1 - 2p)(1-\rho^2)^2\,.
\end{equation}
The profile function is
$\phi(y)=\frac{1}{2}(1-2p)\big(1+(y/(1-2p))^2\big)$ for $|y| \leq 1-2p$ and 
$\phi(y) = |y|$ for $|y| \geq 1-2p$. For the
totally asymmetric case, $p=0$, the limit (\ref{18}) has been established before by Johansson
\cite{J}.  The PushTASEP falls also under our scheme with a proof by Borodin and Ferrari \cite{BF}.
The scaling theory is further confirmed for growth models different from single-step, to mention the discrete time TASEP \cite{J}, the
polynuclear growth model \cite{PS01}, and the KPZ equation \cite{ACQ10,SS}.

The theory of Macdonald process \cite{BC} has brought a $q$-deformed
version of the TASEP in focus. For us here it is a further example
for which the non-universal constants can be computed. For the rate
in (\ref{9}) we set
\begin{equation}\label{19b}
   c_{j,j+1} (\eta)=\tfrac{1}{4}(1-\eta_j)(1+\eta_{j+1}) g((n^-_{j+1}
(\eta))\,,
\end{equation}
where $n^-_{j}$ is the number of consecutive $-$ slopes to the left of
site $j$. $g(0)=0$, $g(j)>0$ for $j > 0$, and $g$ increases at most linearly. The
$q$-TASEP is the special case where $g(j)=1-q^j$, $0\leq q<1$, with
the TASEP recovered in the limit $q\to 0$. The slope system maps
onto the totally asymmetric zero range process for length of
consecutive gaps between $+$ slope, denoted by $Y_j$,
$j\in\mathbb{Z}$, $Y_j = 0,1...$\,. In the stationary measure the $Y_j$'s are i.i.d.
and
\begin{equation}\label{19c}
\mathbb{P}(Y_0=k)=\left\{
            \begin{array}{ll}
              Z(\alpha)^{-1}\,, & {k=0,} \\
               Z(\alpha)^{-1}\big(\prod^k_{j=1} g(j)\big)^{-1}\alpha^k\,,
 & {k=1,2,\ldots,}
            \end{array}
          \right.
\end{equation}
where
\begin{equation}\label{19d}
Z(\alpha)=1+\sum^\infty_{k=1}\big(\prod^k_{j=1}
g(j)\big)^{-1}\alpha^k
\end{equation}
and $\alpha>0$ such that $Z(\alpha)<\infty$. The translation
invariant, time stationary measures for the $\eta(t)$-process with rates
(\ref{19b}) are stationary renewal processes on $\mathbb Z$ with renewal distribution
(\ref{19c}), (\ref{19d}).

The coefficient $A$ can be computed from
\begin{equation}\label{19e}
   \lim_{N\to\infty} \frac{1}{N}\log \big\langle \exp \big[\lambda
\sum^N_{j=1} \eta_j \big]\big\rangle_\alpha =r(\lambda)\,,
\end{equation}
where the average is over the stationary renewal process with
parameter $\alpha$. Then $\rho= r'(\lambda)$ and $A=r''(\lambda)$ at
$\lambda=0$. The rate function $r$ is implicitly determined by
\begin{equation}\label{19f}
 r(\lambda)=-\lambda-\log z(\lambda)\,,\quad \frac{1}{Z(\alpha)} z(\lambda)
Z(\alpha z(\lambda)) \mathrm{e}^{2\lambda}=1\,.
\end{equation}
$\rho$, $A$ are computed by successive differentiations. The result
is best expressed through $G(\alpha)=\log Z(\alpha)$. Then
\begin{equation}\label{19g}
   \tfrac{1}{2}(1+\rho)=(1+\alpha G')^{-1}\,,
\end{equation}
\begin{equation}\label{19h}
   A= 4(1+\alpha G')^{-3} \alpha (\alpha G')' =-\alpha (1+\rho)
\frac{d\rho}{d\alpha}\,.
\end{equation}
The average current is given by
\begin{equation}\label{19i}
 {\sf  j}(\rho)=-2 \langle c_{0,1} \rangle_\alpha\,,\quad {\sf j}=-\alpha
(1+\rho)\,.
\end{equation}
One can use (\ref{19i}) together with (\ref{19g}) to work out $\lambda$. But no particularly
illuminating formula results for the combination $\lambda A^2$.

In addition to (\ref{18}) there is a second scale, which will play
no role here, but should be mentioned. Instead of the height
statistics at the single point $\lfloor yt\rfloor$ one could
consider, for example, the joint distribution of $h(j_1,t)$,
$h(j_2,t)$, referred to as transverse correlations. The transverse scale
tells us at which separation $|j_1-j_2|$ there are nontrivial
correlations in the limit $t\to\infty$. The KPZ scaling theory
asserts that this scale is
\begin{equation}\label{19a}
   (2\lambda^2 A t^2)^{1/3}\,.
\end{equation}
The factor $2$ comes from the requirement that the limit joint
distribution is the two-point distribution of the Airy process.
Corresponding predictions hold for the multi-point statistics. Also
when considering the two-point function of the stationary $\eta(t)$
process, $\mathbb{E}\big(\eta_0(0)
\eta_j(t)\big)-\mathbb{E}\big(\eta_0(0)\big)^2$, up to a shift linear in $t$, $j$ has to vary on
the scale of (\ref{19a}) to have a nontrivial limit as $t\to\infty$.
The scale (\ref{19a}) is confirmed for the PNG \cite{PS04}, TASEP \cite{C}, and PushTASEP \cite{BF} two-point function in case of step initial conditions, for the stationary TASEP \cite{FS05} and stationary KPZ equation \cite{IS},
and for TASEP and PNG \cite{BFPS06,BFPS08} in case of flat initial conditions.


\section{The semi-discrete directed polymer model}\label{sec3}
 \setcounter{equation}{0}

Our starting point is a very particular discretization of the
stochastic heat equation as
\begin{equation}\label{20}
   d Z_j=Z_{j-1}dt + Z_j d b_j\,.
\end{equation}
Here $j\in\mathbb{Z}$, $t\geq 0$, and $\{b_j(t),j\in\mathbb{Z}\}$ is
a collection of independent standard Brownian motions. The analogue
of the wedge initial condition is
\begin{equation}\label{21}
   Z_j(0)= \delta_{j,0}\,.
\end{equation}
Hence $Z_j(t)=0$ for $j<0$ and
\begin{eqnarray}\label{21a}
&&\hspace{-20pt} d Z_j=Z_{j-1}dt + Z_j db_j\,,\quad j=1,2,\ldots\,,
\nonumber\\
&&\hspace{-20pt} d Z_0= Z_0db_0\,.
\end{eqnarray}
Let us introduce the totally asymmetric random walk, $w(t)$, on
$\mathbb{Z}$ moving with rate 1 to the right. Denoting by
$\mathbb{E}_0$ expectation for $w(t)$ with $w(0)=0$, one can
represent
\begin{equation}\label{22}
Z_j(t)=\mathbb{E}_0\big(\exp\big[\int^t_0 db_{w(s)}(s)\big]
\delta_{w(t),j}\big)\mathrm{e}^t\,.
\end{equation}
$Z_j(t)$ is the random partition function of the directed polymer
$w(t)$, length $t$, endpoints 0 and $j$, in the random potential $d
b_j(s)/ds$. This model was first introduced by O'Connell and Yor \cite{OY},
see also \cite{MO,O}. In the zero temperature limit one would maximize over 
the term in the exponential at fixed $\{b_j(s)\}$ and fixed end points, see \cite{GW} for an early study. The statistics of the maximizer is closely related to GUE and Dyson's Brownian motion \cite{B}.

The height corresponds to the random free energy and we set
\begin{equation}\label{23}
   h_j(t)=\log Z_j (t)\,,\quad j\geq 0\,,\;t>0\,.
\end{equation}
$h_j$ is the solution to
\begin{equation}\label{24}
 d h_j=\mathrm{e}^{h_{j-1}-h_j} dt+db_j
\end{equation}
and the slope $u_j=h_{j+1}-h_j$ is governed by
\begin{equation}\label{25}
  d u_j=\big( \mathrm{e}^{-u_j}-\mathrm{e}^{-u_{j-1}}\big)dt +
db_{j+1}-db_j\,,\quad j=1,2,\ldots\,,
\end{equation}
\begin{equation}\label{25a}
du_0= \mathrm{e}^{-u_0}+ db_1-db_0\,.
\end{equation}
The slope $u_j(t)$ is locally conserved.

Somewhat unexpectedly, one can still find the stationary and
translation invariant measures for the interacting diffusions
(\ref{25}) on the lattice $\mathbb{Z}$. They are labeled by a
parameter $r>0$ and are of product form. The single site measure is a double exponential of the form
\begin{equation}\label{25b}
  \mu_r(dx) = \Gamma(r)^{-1}\mathrm{e}^{-\mathrm{e}^{-x}} \mathrm{e}^{-rx}
dx\,,\quad r>0\,.
\end{equation}
Averages with respect to $\mu_r$ are denoted by $\langle
\cdot\rangle_r$. The parameters of the scaling theory are now easily
computed. We find
\begin{equation}\label{26}
\rho=\langle u_0\rangle_r=-\psi (r)
\end{equation}
with $\psi=\Gamma'/\Gamma$ the Digamma function on
$\mathbb{R}^+$. Note that $\psi'>0$, $\psi''<0$, and $\rho$
ranges over $\mathbb{R}$. From (\ref{25}) the random current is
$-\mathrm{e}^{-u_j}dt -db_{j+1}$ and hence the average current
\begin{equation}\label{27}
   \mathsf{j}=-\langle\mathrm{e}^{-u_0}\rangle_r=-r\,.
\end{equation}
Finally
\begin{equation}\label{28}
   A(r)=\langle u^2_0\rangle_r -\langle
u_0\rangle^2_r=\psi'(r)\,.
\end{equation}

Since the initial conditions force $\phi$ to be convex down, the signs from Section 2 are reversed. 
In particular,  the sup in (\ref{15}) is
replaced by the inf and
\begin{equation}\label{29}
  \phi(y)= \inf_{\rho\in\mathbb{R}}\big(-y\rho-{\sf j}(-\rho)\big)\,,\quad
y\geq 0\,.
\end{equation}
$\phi(0) = 0$, $\phi''<0$, and $\phi$ has a single strictly positive maximum
before dropping to $-\infty$ as $y\to\infty$. Thus $t\phi(y/t)$ reproduces the singular initial conditions for
(\ref{24}) as $t\to 0$.
Moriarty and O'Connell \cite{MO} prove that
\begin{equation}\label{30}
\lim_{N\to\infty} \frac{1}{N}h_N (\kappa N)=f(\kappa)
\end{equation}
with
\begin{equation}\label{31}
f(\kappa)=\inf_{s\geq 0}\big(\kappa s-\psi(s)\big)\,.
\end{equation}
The scaling theory claims that
\begin{equation}\label{32}
\lim_{t\to\infty} \frac{1}{t}h_{\lfloor yt\rfloor}
(t)=\phi(y)\,,\quad y>0\,.
\end{equation}
Hence, using that $\psi'>0$ and $y\kappa=1$, we have
\begin{equation}\label{34}
 \phi(y)=\frac{1}{\kappa} f(\kappa) =\inf_{s\geq 0}\big(s-y\psi(s)\big)
 =\inf_{\tilde{s}\in\mathbb{R}}\big(\psi^{-1}
(\tilde{s})-y\tilde{s}\big)\,,
\end{equation}
in agreement with (\ref{29}).

The asymptotic analysis of the height fluctuations is due to Borodin
and Corwin with the result\medskip\\
\textbf{Theorem} (5.2.12 of \cite{BC}). \textit{There exists a
$\kappa^\ast$ such for $0<\kappa^\ast<\kappa$ it holds}
\begin{equation}\label{35}
 \lim_{n\to\infty}\mathbb{P}\big(h_n(\kappa n) - n f(\kappa)\leq
(-2f''(\kappa))^{-1/3} n^{1/3}
s\big)=F_\mathrm{GUE}(s)\,.\medskip
\end{equation}

According to (\ref{27}),  $\lambda = - {\sf j}'' > 0$. Hence in (\ref{18}) $-\lambda$ is replaced by $\lambda$ and the sign to the right of $\leq $ is $+$. To see whether with these changes the scaling theory is confirmed, we start from
\begin{equation}\label{36}
h_{\lfloor yt\rfloor}(t)= t\phi(y)+ (\tfrac{1}{2}\lambda A^2 t)^{1/3}
\xi_\mathrm{TW}
\end{equation}
with $\xi_\mathrm{TW}$ a GUE Tracy-Widom distributed random
variable, hence
\begin{equation}\label{37}
h_n(\kappa n)= \kappa n \phi(\kappa^{-1})+
(\tfrac{1}{2}\lambda A^2 \kappa n)^{1/3} \xi_\mathrm{TW}\,.
\end{equation}
 Now $\rho=-\psi(r(\rho))$ is differentiated as
\begin{equation}\label{38}
1= -\psi' r'\,,\quad 0= \psi'' (r')^2 +\psi' r''\,.
\end{equation}
Since $-\lambda={\sf j}''(\rho)=-r''(\rho)$ and $A(r)=\psi'(r)$, one
has
\begin{equation}\label{38a}
 \lambda A^2=\psi''r'\,.
\end{equation}
Since $y={\sf j}'(\rho)=-r'(\rho)$ and $y\kappa=1$, we conclude
\begin{equation}\label{39}
 \lambda A^2\kappa=-\psi''
\end{equation}
and, since $f$ is the Legendre transform of $\psi$,
\begin{equation}\label{40}
 \lambda A^2\kappa =-\frac{1}{f''}\,,
\end{equation}
in agreement with (\ref{35}).


\section{Conclusion}\label{sec4}
 \setcounter{equation}{0}

The KPZ scaling theory makes a prediction on the non-universal
coefficients for models in the KPZ class and has been confirmed for
PASEP, discrete TASEP, and PNG. We add to this list the semi-discrete
directed polymer. The corresponding stochastic ``particle'' model is
a system of diffusions, $u_j(t)$, with nearest neighbor interactions
such that the sums $\sum_{j} u_j(t)$ are locally conserved. This model
has a   flavor rather distinct from driven lattice gases. Still the
long time asymptotics in all models is the Tracy-Widom statistics.

\end{document}